# RESONANT ION RADIATION SCATTERING AND THE INTEGRATED ATOMIC CROSS-SECTION AS APPLIED TO BINARY STAR SHOCK FRONTS.

Raymond J. Pfeiffer
Professor Emeritus of Physics and Astronomy
Department of Physics, The College of New Jersey, USA
Pfeiffer@tcnj.edu

## ABSTRACT

The current literature is rather vague regarding how to calculate the exact numerical value of the resonant ion scattering cross-section that should be used for a specific bandpass of finite width. Such a value was needed in order to calculate the ion and mass densities in the shock fronts of hot, close binary star systems. This was done based on a modeling of ultraviolet wind-line profiles, using *IUE* spectra. Therefore, a numerical integration has been carried out, in wavelength-space, of the exact expression for the cross-section over two band-passes of astrophysical interest. The exact expression employed was that derived from a solution of the Abraham-Lorentz equation. The numerical results depend on the resonant wavelength, which is taken to be at the center of the bandpass. Most texts on the subject derive an expression for the scattering cross-section in frequency-space, based on the assumption that the radiation reaction term in the Abraham-Lorentz equation may be approximated by a resistive term. The integral of this cross-section over the entire spectrum is independent of the resonant frequency, except for the transition probability. This has limited practical use when dealing with fluxes measured in a bandpass of finite width expressed in wavelength units and scattering is the only mechanism for producing the observed fluxes. Such is the case when dealing with the low densities encountered in stellar winds and shock fronts.

Integrated cross-sections that depend on the resonant wavelength are used to determine the number and mass densities of C IV and N V ions in the shock fronts found in some hot, eclipsing binary star systems, for which several *IUE* spectra have been obtained over a Keplerian orbital period. This then leans to a determination of the mass density and total mass in the shock once the volume of the shock is determined.

## I. INTRODUCTION

This study was prompted by an undertaking to model the UV light curves for band-passes observed with the *IUE* satellite telescope, now retired, encompassing a wind-line profile for hot, eclipsing binary stars. To accomplish this in part, it was necessary to calculate the amount of radiation that is scattered towards the observer, in a specified bandpass, by certain ions in a given volume element of a shock front formed by the colliding winds of the stars. In so doing, one needed to determine the product of the ion density and the atomic scattering cross-section. In turn, the number density of ions, such as C IV and N V, in the shock could be found, if the atomic scattering cross-section were known. The problem then became to find the value of the scattering cross-section, $\sigma_{scat}(\Delta\lambda)$ for the appropriate bandpass, $\Delta\lambda$. One would think that the theory of atomic scattering that is given in texts would be clear on this matter, but this is not the case. Most authors develop the theory of atomic scattering in frequency space rather than wavelength space. This is because angular frequency, $\omega$, is a natural parameter that appears in the harmonic oscillator equation.

That is, ω is related to the ratio of the elastic restoring force constant to the mass of the electron, more specifically, $\omega^2 = k/m_e$. One purpose of this paper is to help clarify this matter when dealing with spectrophotometric data that employs wavelengths. The other is to compute the ion and mass densities in the shock fronts that have been previously modeled for such binary star systems such as EM Carinae (Pfeiffer & Stickland. 2004) and HD159176 (Pfeiffer *et al.* 1997). This in turn leads to the mass in the shock front once the volume is determined.

## 2. RESONANT ION SCATTERING FROM A STATIC SHOCK FRONT

Let $F_j(\lambda)$ be the unpolarized, monochromatic flux at wavelength $\lambda$, emanating from the *j-th* element of surface area of the photosphere (surface) of a star and incident on an ion located in the shock front of a binary star system. Flux here is defined as the radiant energy per unit area per unit time. The total monochromatic power scattered in all directions by that ion is (Stone, 1963, p. 343):

$$P_{scat}(\lambda) = \sigma_{scat}(\lambda) F_j(\lambda). \tag{1}$$

An equivalent equation is (4-31) in Mihalas (1978). Actually, equation (1) defines $\sigma_{scat}(\lambda)$, the atomic scattering cross-section at wavelength $\lambda$. If one assumes that the incident beam is unpolarized, the power, $\Delta P_i$ in bandpass $\Delta\lambda$ that is scattered by a volume element $\Delta V_i$ with ion density $n$, into solid angle $\Delta\Omega$ by angle $\chi_i$ is:

$$\Delta P_i(\Delta\lambda) = \sigma_{scat}(\Delta\lambda)\left[\frac{3}{16\pi}(1 + \cos^2\chi_i)\right] n\Delta V_i F_{ij}(\Delta\lambda)d\Omega. \tag{2}$$

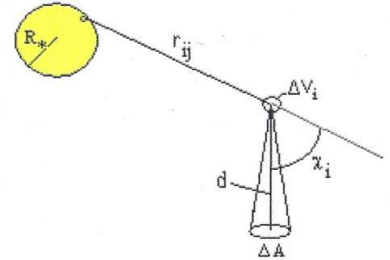

Fig. 1. The geometry for scattering of stellar radiation from a small volume of ions.

See Fig. 1. The factor in the square brackets is the normalized, angular distribution factor for scattering. The factor $F_{ij}(\Delta\lambda)$ is the flux emanating from the *j-th* surface element of a stellar photosphere within bandpass $\Delta\lambda$ and incident on the *i-th* volume element of the shock. The volume element of the shock, $\Delta V_i$, is located at a distance $r_{ij}$ from surface element $j$ of a star, which has radius $R_*$. Equation (2) is based on the development given by Stone (1963), who uses irradiance, I, instead of flux but otherwise presents a rather clear presentation of scattering theory not found in most sources. If $\Delta A$ is the area of a detector at distance $d$ from the volume element $\Delta V_i$ of the shock front, then $\Delta\Omega = \Delta A/d^2$, and the flux falling on the detector from that volume element is:

$$\Delta F_i(\Delta\lambda, d) = \Delta P_i(\Delta\lambda)/\Delta A = \sigma_{scat}(\Delta\lambda)\left[\frac{3}{16\pi}(1 + \cos^2\chi_i)\right] n\Delta V_i F_{ij}(\Delta\lambda)/d^2. \tag{3}$$

To simplify this expression for further discussion, let us define the scattering factor:

$$S_i = \sigma_{scat}(\Delta\lambda)\left[\frac{3}{16\pi}(1 + \cos^2\chi_i)\right]$$

Then (3) simply becomes:

$$\Delta F_i(\Delta\lambda, d) = S_i n\Delta V_i F_{ij}(\Delta\lambda)/d^2 \tag{4}$$

Now let $F_{*j}(\Delta\lambda, R_*)$ be the value of the flux emanating from a surface element j of the star, at the surface of the star. Then, $F_{ij}(\Delta\lambda, r_{ij}) = F_{*j}(\Delta\lambda, R_*)\frac{R_*^2}{r_{ij}^2}$. That is $F_{*j}(\Delta\lambda)$ under goes a radiation dilution factor before reaching the shock volume element. Then (4) becomes:

$$\Delta F_i(\Delta\lambda, d) = \mathcal{S}_i \left[ F_{*j}(\Delta\lambda, R_*) \frac{R_*^2}{r_{ij}^2} \right] n \Delta V_i / d^2 \tag{5}$$

In addition, the flux undergoes an additional attenuation factor $\alpha(\Delta\lambda) = e^{-\tau}$ that is caused by the intervening wind envelopes, of which there is one for each star. Here $\tau$ is the mean optical depth along the direction $r_{ij}$. The latter quantity must be determined by a model of the binary star system. Now (5) becomes:

$$\Delta F_i(\Delta\lambda, d) = \mathcal{S}_i \left[ F_{*j}(\Delta\lambda, R_*) \frac{R_*^2}{r_{ij}^2} \right] \alpha(\Delta\lambda) n \Delta V_i / d^2 \tag{6}$$

Furthermore, $F_{*j}(\Delta\lambda)$ is some fraction of the total flux $F_*(\Delta\lambda, R_*)$ emanating from the stellar surface. That is $F_{*j}(\Delta\lambda) = F_*(\Delta\lambda, R_*) p_j$, were $p_j$ is the value of that fraction, that is, $p_j$ is a surface flux partitioning or distribution factor. We may now write (6) as:

$$\Delta F_i(\Delta\lambda, d) = \mathcal{S}_i \left[ F_*(\Delta\lambda, R_*) p_j \frac{R_*^2}{r_{ij}^2} \right] \alpha(\Delta\lambda) n \Delta V_i / d^2 \tag{7}$$

This is essentially the derivation of equation (3) in Pfeiffer *et al.* (2004), but with a slightly different notation and considering only one of the stars in the binary. However, the above derivation is not presented there. Furthermore, the parameter $F_{ijk}$ in equation (3) of the above reference, where *k* indexes a specific star in the binary, is the equivalent of $F_{ij}$ in equation (3) of this paper.

Now the the total observed flux in any bandpass $\Delta\lambda$, $F_{obs}(\Delta\lambda, d)$, is known for each of the several *IUE* spectrophotometric images that have been acquired over a Keplerian orbital cycle. The contribution to this total flux from the shock is found by integrating (7) over the volume of the shock, which is assumed to be static. To accomplish this, one needs to determine $p_j$ for each surface element of each star as seen by the shock volume element $\Delta V_i$. For the photospheres of the stars, $p_j$ includes limb darkening, polar brightening (also called gravity darkening when referring to the stellar equatorial latitudes) and Lambert's law. The integration is itself a very complex geometric procedure, since the scattering angle $\chi_i$ varies from one volume element in the shock to another. In practice, this angle is found from the inner vector product of the position vector of the volume element relative to an element of area of a photosphere and the unit directional vector of the observer. See Fig. 2, taken from Pfeiffer & Stickland (2004), which depicts the model developed for the system EM Carinae. This model results from a program written by the author. Each dot in the diagram is a volume element of the winds, the surface elements of the stars are crosses that intersect appearing to form a network of patches and the shock front volume elements are the diamonds. Note that the stars are tidally distorted.

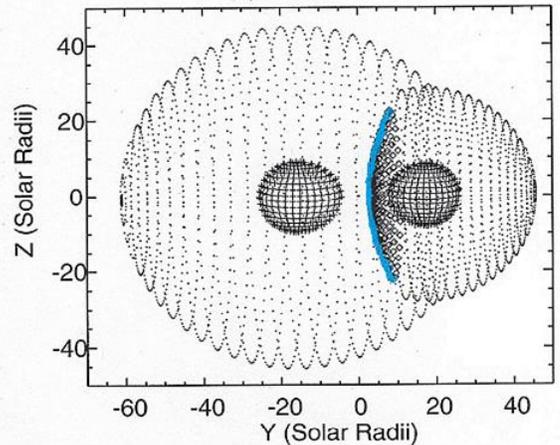

Fig. 2. The grid of points used to model the EM Carinae binary star system as seen in the plane of the sky at orbital phase 0.264, when the larger star is approaching the observer. The wind envelopes (dots) are reresented as spheres truncated where they intersect. The shock front volume elements conists of the grid of diamonds.

Also, the factor $\alpha(\Delta\lambda)$ must be determined by taking into account the attenuation of the photospheric flux as it traverses the wind envelope before encountering the volume element in the shock and again after it is scattered and traverses the wind or winds again in the direction of the observer. To accomplish this, one needs to model the opacity in the wind as a function of path length.

Once the integration over the volume of the shock front is done, the resulting flux from the shock is added to the fluxes emanating from the photospheres and winds of the stars, taking into account eclipsing effects on all parts of the system. However, all these details are not our concern here, but may be found in either Pfeiffer *et al.*(1994) or Pfeiffer & Stickland (2004). When the scattered flux contributed by the shock to the total observed systemic flux for the bandpass is correctly modeled, the number density of the ions in the shock may be found, provided the value of $\sigma_{scat}(\Delta\lambda)$ is known. This is the matter we now address.

## 3. THE INTEGRATED SCATTERING CROSS-SECTION

We first begin the discussion in frequency space, $\omega = 2\pi\nu$, since this is what is most commonly found in the literature. Later, we shall transform to wavelength space. Now the scattering cross-section is the following function of $\omega$:

$$\sigma(\omega) = \sigma_T \left[\frac{\omega^4}{(\omega^2 - \omega_0^2)^2 + (\gamma\omega^3/\omega_0^2)^2}\right]$$

The function in the square brackets is often referred to as the scattering coefficient. Here $\sigma_T$ is the Thomson cross-section equal to 6.6525 x $10^{-25}$ cm$^2$, $\omega = 2\pi f = 2\pi c/\lambda$, , $\gamma = 2e^2\omega_o^2/(3m_e c^3)$ is the classical radiative decay constant, $e$ is the electrical charge on the electron, $m_e$ is the mass of the electron, c is the speed of light, and $\omega_o$ is the resonant frequency of an atomic transition.

The total or integrated scattering cross-section, $\sigma_I$, is obtained by integrating $\sigma(\omega)$ over all frequencies from zero to infinity, viz.,

$$\sigma_I = \sigma_T \int_0^\infty \frac{\omega^4 d\omega}{(\omega^2 - \omega_0^2)^2 + (\gamma\omega^3/\omega_0^2)^2} \tag{8}$$

Note that the integrand in (8) is dimensionless and therefore the dimensions on $\sigma_I$ are those of $\sigma_T d\omega$, that is cm$^2$-Hz. The above expression is equation (14-15) in Stone (1963). Similar expressions are equations (4-81) in Aller (1963), (17.63) in Jackson (1974), (4-34) in Mihalas (1978), (6.71) in Padmanabhan (2000) and (3.62) in Rybicki & Lightman(1979). Some authors use the symbol $\Gamma$ instead of $\gamma$.

One would suppose that all these authors give the same expression, but this is not the case. The disagreement concerns the 2nd term in the denominator of the integrand in (8). The expression for this term depends on what assumptions are made in the solution of the Abraham-Lorentz equation. Let us define this term to be $u$. Following Jackson, and Rybicki & Lightman, we introduce the characteristic time, $\tau = 2e^2/(3m_e c^3)$, so that $\gamma$ or $\Gamma = \tau\omega_o^2$. Then for Mihalas, $u$ is $\gamma^2\omega^2$ or $u = \tau^2\omega_o^4\omega^2$. Rybicki & Lightman give $u = \tau^2\omega_o^6$, which is suspected to be a misprint, and should have been $u = \tau^2\omega^6$. Stone gives $u = (\gamma\omega^3/\omega_o^2)^2$ or $u = \tau^2\omega^6$, and Padmanabhan gives $u = \tau^2\omega_o^4\omega^2$ in agreement with Mihalas. The interpretation of Jackson's equation (17.63) is that it is in agreement with the expression given by Stone. This expression is obtained by taking the resistive term in the Abraham-Lorentz equation to be zero and keeping the radiation reaction term. That is, there is no energy loss by the harmonic oscillator to other effects (such as atomic collisions) other than quantum radiation (radiative scattering) and that one is dealing with a steady state solution. Hence, $\Gamma' = 0$ in $\Gamma_t = \Gamma' + \Gamma$ (Jackson, equation 17.61). Stone and Rybicki & Lightman do not approximate the radiation reaction term by using a resistive term, but Aller, Mihalas and Padmanabhan do. This leads to the differences in the solutions among the above authors. Of course, quantum mechanics introduces the oscillator strength or transition probability, $f_{ij}$, so that

$$\sigma_{scat} = \sigma_I f_{ij} \tag{9}$$

For the following discussion, only the classical case shall be considered, since the final results need only to be multiplied by the transition probability. The most transparent derivation of (8) is that given by Stone (Chapters 12 and 14) and the most rigorous by Jackson (Chapter 17). Hence, (8) is the correct and most rigorous one when one is dealing with measured fluxes that are the result of just scattering by an atom or ion, and not scattering plus absorption (total extinction, as defined by most authors). Aller's exposition on this matter which leads to his equation (4-81) is somewhat confusing. On page 169, he states that the absorptivity, denoted by $\nu$, as given by equation (4-73) was derived from a consideration of light scattering.

In reality, the derivation of (4-73) was carried out on the basis of pure absorption using a resistive term in the equation of harmonic motion, (4-61), rather than a radiation reaction term.

The usual assumption made by some authors is that the profile of the scattering coefficient is so sharply peaked and the term **u** so very small, its exact value is not important. The replacement of the radiation reaction term by only a resistive term appears to be done in order to achieve and expression for σ that may be readily integrated. As it will be shown, this turns out to be numerically valid for adequately wide band-passes. Since the advent of high-speed computers, it appears that no one has attempted a numerical integration of (8) for finite band-passes.

However, even (8) is somewhat heuristic and perhaps misleading, for it is not the fundamental expression that defines the classical scattering cross-section, $\sigma_{scat}$. The atomic cross-section for scattering has no independent meaning outside of equation (1). That is, $P_{scat}(\Delta\lambda)$, or $P_{scat}(\Delta\omega)$, has a value that is specific for a particular bandpass and is to be found only by convolving the expression for the integrand in (8), namely,

$$\left[\frac{\omega^4}{(\omega^2 - \omega_o^2)^2 + (\gamma\omega^3/\omega_o^2)^2}\right]$$

with the flux $F(\omega)$ when integrating over the bandpass of interest. In other words, when carrying out the integration indicated in (8), it is usually assumed that the radiation field is constant over all frequencies or wavelengths, which is never the case. Aller (1953, page 166) is the only author who warns about this, saying that $F_v$ should not vary appreciably with frequency near the resonant frequency. Additionally, most authors assume that σ is such a sharply peaked function, the contribution to the integration over frequency is negligible except in the region immediately near the resonant wavelength or frequency, even when one integrates to infinity. It was decided to thoroughly investigate this assumption and find out how wide the bandpass must be for this to be true, using the exact expression, equation (8), for the integrated cross-section in the case of pure scattering.

Even if (8) is assumed valid, independent of $F(\omega)$, the integral has no analytical solution. Hence, the need to use the alternative version of (8) that is derived using the resistive term in the Abraham-Lorentz equation in place of the radiation reaction term. In addition, it is assumed that $\omega \approx \omega_o$ and $\omega^2 - \omega_o^2$ may be approximated by $2\omega(\omega - \omega_o)$ (Aller, p. 164 & 167; Mihalas, p. 83; Padmanabhan, p. 267; Rybicki & Lightman, p. 101). With these simplifying assumptions, and transforming variables from ω to $v$, where $\omega = 2\pi v$, the integration leads to:

$$\sigma_I = \frac{\pi e^2}{m_e c} \quad (cm^2/sec) \tag{10}$$

when integrating from 0 to +∞. The charge on the electron, e, is in esu units. There is an additional factor of 2π when integrating using ω as the variable of integration (Jackson, equation 17-73; Rybicki and Lightman, equation 3-65a). It is interesting to note that (10) is independent of the resonant wavelength or frequency. Jackson (page 805) further enlightens this matter by stating that (10) is obtained by neglecting the radiation reaction term and this is equivalent to assuming the width $\Gamma_t$ is independent of the frequency. In other words, the decay time is assumed to be the same for all transitions. Of course, (10) is of little use unless it is used in conjunction with a broadening function that introduces the resonant frequency when calculating line profiles. The most straightforward exposition of this is given by Swihart (1963, p. 131f.) and also by Lamers *et al.* (1999, p. 198f.). However, as is shown below, for all practical band-passes involving pure scattering, σ is dependent on the resonant wavelength, $\lambda_o$, independently of a broadening function.

## 4. THE BANDPASS DEPENDENT VALUE OF $\sigma_{scat}$

It was decided to investigate how good the above simplifying assumptions to obtain (10) are by numerically integrating the exact expression, equation (8), for several band-passes of astrophysical interest.

Additionally, this was done using wavelength units in centimeters instead of frequency. Therefore, to make a comparison, it was needed to convert frequency units to wavelength units in centimeters. In addition, it was needed to convert the numerical value for $\sigma_I$ from (10) into wavelength space, which is 2.65376 x 10$^{-1}$ cm$^3$. (Values for the constants in (10), with a precision to at least 7 places, were taken from Cox (2000).) Such a conversion may be computed from

$$\sigma(\lambda) = \sigma_I f \lambda_o^2 /c \quad (cm^3). \tag{11}$$

The latter is given by Aller (1963, p. 302), when using wavelength units for computing the equivalent widths of broadened lines. Also see the discussion by Unsöld *et al.* (2001, p. 203f.) regarding line broadening. It comes about by changing the variable of integration from $d\nu$ to $d\lambda$, since $(\lambda^2/c)d\nu = |d\lambda|$ and $\lambda$ is evaluated at $\lambda_o$.

Now, for one of the C IV resonant wavelengths taken from Striganov & Sventitskii (1968), equation (11) gives $\sigma(\lambda)$ = 2.12172 x 10$^{-20}$ cm$^2$-cm, (uncertainty in the last place) with *f*, the transition probability, taken to be unity. The results from our numerical integration shall be compared with this value. Converting (8) to wavelength units and then integrating over a finite bandpass centered on the resonant wavelength we get:

$$\sigma_{\Delta\lambda} = \sigma_T \int_{\lambda_i}^{\lambda_f} \frac{d\lambda}{(1 - \frac{\lambda^2}{\lambda_o^2})^2 + \eta/\lambda^2} \quad (cm^3) \tag{12}$$

In (12), $\eta$ is the constant **$4\pi^2 c^2 \tau^2 = 16\pi^2 e^4/9m_e^2 c^4$**, which has the value 1.39303 x 10$^{-24}$ cm$^2$. Again, values for the constants were taken from Cox (2000). Notice that the denominator in (12) is dimensionless and that the dimension of the integral is just cm. The dimensions of $\sigma_{\Delta\lambda}$ are then cm$^3$.

Values for (12) were calculated by writing a FORTRAN program that numerical integrates (12) by the method of quadratures with unit weights on a Sun Blade 150 work station. In order to avoid inconsistent results, σ must be evaluated at $\lambda_o$ and multiplied by $d\lambda$ in one of the steps. This was accomplished by starting the integration at $\lambda_o$ and then integrating from there to shorter and longer wavelengths separately, since the function is not symmetrical about $\lambda_o$. Numerical values for the step in the integration, $d\lambda$, were used ranging from 1 x 10$^{-8}$ cm to 1 x 10$^{-16}$ cm. It was found that the integral converges to essential the same number, significant to 6 places, after $d\lambda$ is reduced to 1.0 x 10$^{-13}$ cm. However, the value for $\sigma\Delta\lambda$ increases slightly with increasing bandwidth. The results for one of the resonant wavelengths for C IV, $\lambda_o$ = 1548.185Å (Striganov *et al.*, 1968) are listed in Table 1. Uncertainty there is in the last decimal place. The last entry in Table 1 is the result using equation (11). Notice how the results converge to the value given by (11) as the bandwidth increases. These values differ from the value obtained from equation (11) in the 3rd decimal place for the smallest band-pass and in the 4th place for band-passes less than 10 Å in total width. Therefore, the simplifying assumptions that the radiation reaction term in the Abraham-Lorentz equation may be approximated by a resistive term and that $\omega^2 - \omega_o^2$ may be approximated by $2\omega(\omega - \omega_o)$ are very good ones and testify to how sharply peaked the profile of the scattering coefficient is.

Table 1
Values of the integrated scattering cross-section centered on the resonant wavelength 1548.185 Å for the C IV ion for several different bandpasses. The transition probability is set equal to 1.0.

| $\Delta\lambda$ (Å) | Bandpass (Å) | $\sigma$ (cm$^3$) |
|---|---|---|
| 1.0 | 1547.685 – 1548.685 | 2.12156 e-20 |
| 2.0 | 1547.185 – 1549.185 | 2.12164 e-20 |
| 10.0 | 1543.185 – 1553.185 | 2.12170 e-20 |
| 35.0 | 1543.185 – 1553.185 | 2.12171 e-20 |
| | $\sigma_I \lambda_o^2 /c$ | 2.12172 e-20 |

That is, the function for the scattering coefficient is essentially a delta-function. So, when working with wavelength units in cm, one may use (11) for all practical band-passes, but certainly not (10). Of course, this depends on the precision with which one is working.

Integrations over band-passes up to 500 Å wide were also carried out with no change in the result. A 35 Å-wide band-pass was used for the light curve analyzed by Pfeiffer *et al.* (2004). Actually, this bandpass contains a doublet consisting of the above line separated from the other by about 2.6Å.

A similar integration was carried with similar results for a visible bandpass, viz., for the Na I line at 5889.9504 Å with $f$ =1.0. The results also show that the value for the integrated cross-section depends on the resonant wavelength, when no simplifying assumptions are made in the solution of the Abraham-Lorentz equation or in the integration process and without introducing a broadening function.

The result for a 33 Å wide bandpass for one of the components of the N V doublet at 1240.146Å is 3.43 x $10^{-20}$ cm³. This is also in numerical agreement with the value computed from equation (11). The latter bandpass was used for the calculation of the N V density in the shock front of HD159176, which is presented below.

Graphs of $\sigma$ versus $\omega$ are presented by Jackson (1974, pg. 803), Rybicki & Lightman (1979, pg. 10) and Stone (1963, pg. 347) with relative scales. But these figures really do not convey how sharply peaked the function is, though Stone (1963, p. 347) attempts to quantify this. Values of $\sigma$ for various wavelengths centered on the C IV resonant wavelength at 1548.185 Å were computed from $\sigma_I[12m_ec/8\pi^2e^2]$ (Stone, p. 346). These values are given in Table 2, demonstrating this. Even a logarithmic scale on a graph would

Table 2

Computed values for the scattering cross-section at several different wavelengths in the vicinity of the resonat wavelength 1548.185 Å for the Carbon IV ion.

| $\lambda$(Å) | $\sigma$(cm²) |
|---|---|
| 1547.685 | 1.59474e-18 |
| 1548.145 | 2.49103e-16 |
| 1548.155 | 4.42846e-16 |
| 1548.165 | 9.96392e-16 |
| 1548.175 | 3.98544e-15 |
| $\lambda_o$=1548.185 | 1.14442e-10 |

have difficulty showing this, since there are 8 orders of magnitude variation of $\sigma$ over just 0.5 Å.

5. ION DENSITIES AND MASSES IN SOME MODELED SHOCK FRONTS

In addition, the determination of a numerical value for $n\sigma(\Delta\lambda)f$ for its use as an input parameter in the numerical integration of (7) over the volume of a shock front, $n\sigma(\Delta\lambda)f$ must satisfy the relationship that involves the optical depth, $\tau$, of the shock. (Unfortunately, $\tau$ is the symbol also used frequently for characteristic time). That is, when computing the attenuation or eclipsing effects of the shock on the integrated fluxes, within the bandpass $\Delta\lambda$, emanating from the photospheres and winds of either star in the binary, one uses:

$$\tau_{\Delta\lambda} = nft \int \sigma(\lambda)\Phi(\lambda)d\lambda \qquad (13)$$

Optical depth, like $f$, is dimensionless, and so the factors on the left must tally to be so also. Here, $\Phi(\lambda)$ is the usual broadening function. The number density of the ions is assumed to be constant over the thickness of the shock, t. Additionally, the correction for stimulated emission is assumed to be about 1, since shock densities are sufficiently low that most of the ions may be assumed to be in the ground state. Anyway, it is not possible to calculate what this factor should be. Since we are not interested in the details of a line profile, the broadening function may be replaced by simply dividing the result of the integration by the Doppler broadened width of the bandpass, $\Delta\lambda_D$. This is necessary in order that the optical depth be dimensionless. That is, a mean vale for $\tau_{\Delta\lambda}$ may be used for the bandpass:

$$\overline{\tau_{\Delta\lambda}} = nft(\sigma_{\Delta\lambda})/\Delta\lambda_D \qquad (14)$$

The value of the total mass density in a shock may be calculated as follows:

$$\rho_{shk} = (1 + \frac{n_{He}}{n_H} A_{He} + \frac{n_C}{n_H} A_C + \frac{n_N}{n_H} A_N + \frac{n_z}{n_H} A_z) n_H m_H \qquad (15)$$

Here $A$ is the atomic weight, $m_H$ is the mass of the hydrogen atom (and $z$ represents all the other atoms other than helium, carbon, and nitrogen. Since the metal abundances are relatively small, they may all be represented by one term. Values for the abundance ratios relative to hydrogen were taken from Cox (2000) to be 0.090 for helium, 3.63 x 10$^{-4}$ for carbon and 8.80 x 10$^{-3}$ for nitrogen. For the other metals, which are partially ionized to some unknown degree, a weighted mean value has been taken to be 20.0 for $A_z$ and $n_z/n_H$ = 0.0001. The exact values are not significant, since and $n_z/n_H$ is so small. The sum of all the terms in the parentheses in (15) is then 1.498 and it follows that

$$\rho_{shk} = (1.498)[n_m / (n_r/n_H)] m_H. \qquad (16)$$

The number density $n_m$ is that measured for an atom in the shock and $n_r$ is the number density based on the abundance of that atom relative to hydrogen.

### 5.1. The EM Carinae System

The value for $n_{CIV}$ in the shock front was determined by Pfeiffer and Stickland 2004) to be 1560 ions per cm$^3$ and $\tau_{shk}$ is given to be 0.049 for a nominal thickness of the shock to be 2 x 10$^{11}$ cm. If one uses 1560 ions per cm$^3$ for $n_m$ in equation (16), the value for the mass density o f the shock is 1.08 x 10$^{-16}$ g/cm3. This is assuming that all the carbon atoms in the shock are C$_{IV}$ ions.
Now, the shock as modeled for EM Car is located at a distance less than one stellar radius from both the primary and secondary stars. See Figure 2. Typical wind densities, $\rho_w$, in hot stars at photospheric distances of r/R$_*$ = 1 are about 10$^{-13}$ g/cm$^3$ (Lamers *et al.*, p. 251). Therefore, one would expect the density in the shock to be much higher. Additionally, the densities found for the shocks in W-O-type binaries have been modeled by Gayley *et al.* (1996) to be on the order of 10$^{-12}$ g/cm$^3$ or higher. For close binaries, shocks are generally radiatively cooled. In such cases, there may be a factor of 10 to 100 increase in density from the pre-shock wind (Owocki, 2009). So, one would expect $\rho_{shk}$ to be on the order of 10$^{-11}$ g/cm$^3$ or so. So, if the value of $n_{CIV}$ is correct, it cannot be all the carbon atoms in the shock.

Assuming $\rho_{shk}$ = 1.0 x 10$^{-11}$ g/cm3, equation (16) demands that $n_C$ be 1.45 10$^9$ ions per cm$^3$. Hence, the number of C IV is some small fraction of all the carbon atoms in the shock. More than likely, this is the case and that there are other, higher states of ionization such as C V in the shock but not C III. This is because there are no ostensible C III wind-lines noticeable in the *IUE* spectra, such as the doublet near 1428Å or the sextuplet at 1175 Å. Using the value for $n_{CIV}$ to be that mentioned above, the fraction of C IV ions to the total number of carbon atoms is 1560 /1.45 x 10$^9$ = 1.1 10$^{-6}$. Since we do not know the number density for the C V ions, we cannot use the Saha equation to estimate the temperature in the shock, but it must be very high.

Additionally, EM Car is known as an X-ray source (Corcoran, 1996), also indicating that there are very high temperature regions somewhere in the EM Car system, most likely in the shock front. From Fig. 2, the approximate volume of the shock for EM Car may be estimated to be 6.10 x 10$^{36}$ cm$^3$. Using a value equal to 1.0 x10$^{-11}$ for the mass density of the shock in EM Car leads to a mass of 6.10 x 10$^{25}$ grams, or 3 x 10$^{-8}$ solar masses. This may be compared with the mass of the more massive star in the system which is 21 solar masses (Pfeiffer & Stickland 2004).

### 5.2. The HD 159176 System

In the case of HD 159176, Pfeiffer *et al.* (1997) obtained an optical depth equal to 0.93 for the C IV in a modeled shock with nominal thickness 2R$_\odot$. The result for the C IV number density in the shock as reported in that paper is $n = 0.07$ cm$^{-3}$. With this value for the C IV density, the mass density in the shock would be 4.83 x 10$^{-22}$ g per cm$^3$. Again, this is assuming that all the carbon atoms in the shock are C$_{IV}$ ions. Since the system consists of a pair of O6 or O7 spectral type stars, the shock density should be greater in HD 159176 than for EM Car, say at least 5 x 10$^{-11}$ grams per cm$^3$. The exact value is not significant.

Then equation (16) indicates that $n_C$ should be 7.2 10$^9$ atoms per cm$^3$. The number of C IV ions in the shock is then a very small fraction (9.7 x 10$^{-12}$) of all the carbon atoms in the shock. As in the case for EM Car, more than likely, there exist higher states of ionization such as C V in the shock. But since there are rather strong C IV spectral lines in the *IUE* spectra, the temperature is not so high that there are no C IV ions present.

The N V wind line of HD 159176 is strong and allows for a similar analysis as was done for C IV. The above authors obtained an optical depth of 1.01 for the shock, which led to an ion number density $n_{NV}$ = 0.13 cm$^{-3}$. Now the N V line profile is actually a triplet with the wavelength 1240.146 Å as the dominant one. The weighted mean value for the transition probability, $f$, is 0.183. So, equation (11) yields $\sigma_{\Delta\lambda}$ = 2.46 x 10$^{-21}$. Assuming the mass density for the shock front is 5 x 10$^{-11}$ g/ cm$^3$, the number of density of nitrogen atoms would be 1.76 x 10$^{11}$ per cm$^3$. The number of nitrogen atoms is greater than that for carbon simply because the relative abundance for nitrogen is greater. In any event, the number of N V ions in the shock is orders of magnitudes smaller than the total number of nitrogen atoms. The conclusion is that the degree of ionization in the shock front is extremely high. Therefore, a very high temperature environment exists in the shock front that results from the thermalization of the kinetic energy carried by very fast and dense winds as they collide to form the shock front..

The volume of the shock for HD176159 may be computed from the data in Table I of Pfeifer *et al.* (1997) to be 3.8 x 10$^{36}$ cm$^3$. Using the estimated mass density in the shock to be 1.5 x 10$^{-11}$ g/cm$^3$, the mass comprising the shock front is 5.7 x 10$^{25}$ g or 2.9 x 10$^{-8}$ solar mass units. This may be compared with the masses of the stars in the system that are in excess of 40 solar mass units.

### 5.3. The TU Muscae System

The TU Muscae system is a different story, for it contains a slightly over-contact pair of stars (Terrell *et al.*, 2003). A preliminary investigation of the *IUE* spectra for this system by the author (unpublished) has led to a model of the system based on a fitting of ultraviolet light curves and deploying the binary interaction program *BSWI* described in Pfeiffer & Stickland (2004). Fitting the UV light curves was accomplished by developing a program employing techniques similar to those found in Kallrath & Milone (1999). The results of fitting the UV light curves are displayed in Fig. 3. The latter displays three different continuum light curves for the system. The diamonds represent observational data derived from the *IUE* blanketed continuum (1450 - 1490 Å) outside the C IV line profile. Hence, there is little contribution to this light curve from the winds. The star symbols are data points taken from the published light curve of Terrell *et al.* in the *u* bandpass, renormalized to the IUE flux levels. The solid curve is the fit using the program mentioned above that was developed by the author. The primary purpose of achieving such a fit is to test the algorithms so developed, that when they are integrated into the binary interaction program, *BSWI* (Pfeiffer & Stickland, 2004), the latter may be confidently brought to bear upon fitting the light curve of the C IV wind-line profiles.

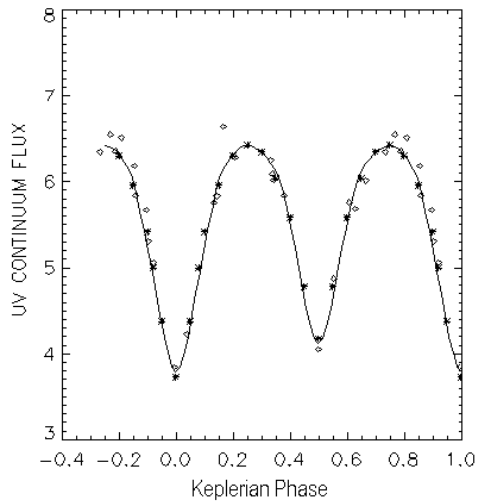

Fig. 3.
The continuum flux is in units of 10$^{-10}$ ergs/cm$^2$/sec.

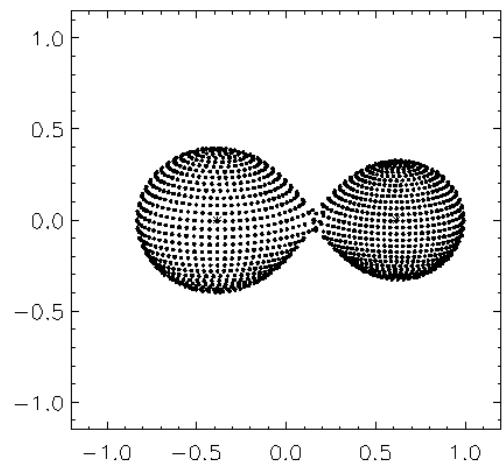

Fig. 4. The TU Muscae system as represented by the algorithms that produced the light curve fit shown in Fig. 3. Other parameters were taken from from Terrell *et al.*

The binary stars, as represented by these algorithms are depicted in Fig. 4. The distance scale used was a convenient one for plotting the data points and the units are not that those of any particular system. The profiles of the two stars were computed using a program employing methods also found in Kallrath & Milone (1999).

The results of the *BSWI* program found no evidence of a shock front in this system. This is probably the result of the over contact configuration of the binary. That is, radiation breaking (Gayle *et al.*, 1996) and the close proximity of the stars to one another does not permit the winds to accelerate to sufficiently high speeds that when they collide, an interaction does not result that would produce a detectable enhancement in the density, namely, a shock front.

## 7. CONCLUSIONS AND RESULTS

The atomic resonant scattering profile is so sharply peaked, that the approximations made in most texts for the solution of the Abraham-Lorentz equation are very good ones to 5 significant figures, when integrating the cross-section over bandpasses greater than 2Å in width. For band-passes narrower than 2 Å, down to 0.50Å, the assumptions leading to (11) are even good to 4 significant figures. One certainly does not want to use the value given by (10) for the cross-section for all bandpasses, since the integrated value of $\sigma_I$ is definitely dependent on the resonant wavelength, independently of a broadening function. However, this matter is not presented uniformly or clearly in most textbooks.

Densities for C IV and N V ions in two binary systems, have been previously published by the author and colleagues. This was accomplished by modeling of the C IV and N V line profiles in *IUE* spectrophometric images, using a verified approximate expression for the integrated scattering cross-sections in wavelength space. Assuming that C IV and N V are the only ions present in the shocks of EM Car and HD159176 leads to mass densities smaller than those found in the literature. This implies that most carbon and nitrogen atoms in the shock fronts are very highly ionized and, hence, very high temperatures environments exist in the shocks. Very high temperatures would certainly result if the wind-collision interactions are very strong. This is in agreement with the wind velocities that have been found to be in excess of 2000 km/sec.

Estimating the total mass densities for the shocks, the fractional abundances of the ions to the total number of ions in the shocks has been calculated. Below are tables summarizing the results for the shock front properties of EM Car and HD 159176. The asterisks on the shock densities mean they are assumed

EM Carinae

| Volum(cm$^3$) | 6.1 | x 10$^{36}$ |
|---|---|---|
| $n_{CIV}$ (ions/cm$^3$) | 1.560 | x 10$^3$ |
| $\sigma_{CIV}$ (cm$^3$) | 2.1217 | x 10$^{-21}$ |
| $\rho_{shk}$* (g/cm$^3$) | 1.0 | x 10$^{-11}$ |
| $n_{CIV}/n_C$ | 1.1 | x 10$^{-6}$ |
| M (g) | 6.1 | x 10$^{25}$ |
| M (M$_\odot$) | 3.0 | x 10$^{-5}$ |

HD 159176

| Volume (cm$^3$) | 3.8 | x 10$^{36}$ |
|---|---|---|
| $n_{CIV}$ (ions/cm$^3$) | 7.0 | x 10$^{-2}$ |
| $\sigma_{CIV}$ (cm$^3$) | 2.1217 | x 10$^{-21}$ |
| $n_{NV}$ | 1.3 | x 10$^{-1}$ |
| $\sigma_N$ (cm$^3$) | 2.4617 | x 10$^{-21}$ |
| $\rho_{shk}$* (g/cm$^3$) | 5.0 | x 10$^{-11}$ |
| $n_{CIV}/n_C$ | 9.7 | x 10$^{-12}$ |
| $n_{NV}/n_N$ | 7.4 | x 10$^{-13}$ |
| M (g) | 5.7 | x 10$^{25}$ |
| M (M$_\odot$) | 2.9 | x 10$^{-22}$ |

values taken from the literature previously cited. It is surprising that the total mass of the shock front in EM Car is greater than for HD 159176, even though the assumed mass density is greater for HD 159176. But this because the shock volume of the later system, as published, is smaller.

It would appear that shock fronts in contact binary systems are not readily detectable by an analysis using the *BSWI* Program.

## ACKNOWLEDGEMENTS

The author wishes to express his gratitude to my colleagues Thulsi Wickramasinghe, Romulo Ochoa and Paul Wiita in the Department of Physics at TCNJ for the many helpful discussions regarding the subject matter of this paper. I also would like to extend my appreciation to Stan Owocki, at the Bartol Research Institute, for the conversations that I have had with him about shock front physics.